# Epitaxial growth of large-gap quantum spin Hall insulator on semiconductor surface

*Miao Zhou[1], Wenmei Ming[1], Zheng Liu[1], Zhengfei Wang[1], Ping Li[1,2], and Feng Liu[1,3*]*

[1] Department of Materials Science and Engineering, University of Utah, UT 84112

[2] School of Physics and Technology, University of Jinan, Jinan, Shangdong, China 250022

[3] Collaborative Innovation Center of Quantum Matter, Beijing, China 100871

*Corresponding author:

Address: Room 304, 122 S. Central Campus Drive, Salt Lake City, UT 84112

Tel: 801-587-7719

Email: fliu@eng.utah.edu

# Abstract

Formation of topological quantum phase on conventional semiconductor surface is of both scientific and technological interest. Here, we demonstrate epitaxial growth of 2D topological insulator, i.e. quantum spin Hall (QSH) state, on Si(111) surface with a large energy gap, based on first-principles calculations. We show that Si(111) surface functionalized with 1/3 monolayer of halogen atoms [Si(111)-$\sqrt{3}\times\sqrt{3}$-X (X=Cl, Br, I)] exhibiting a trigonal superstructure, provides an ideal template for epitaxial growth of heavy metals, such as Bi, which self-assemble into a hexagonal lattice with high kinetic and thermodynamic stability. Most remarkably, the Bi overlayer is 'atomically' bonded to but 'electronically' decoupled from the underlying Si substrate, exhibiting isolated QSH state with an energy gap as large as ~ 0.8 eV. This surprising phenomenon is originated from an intriguing substrate orbital filtering effect, which critically select the orbital composition around the Fermi level leading to different topological phases. Particularly, the substrate-orbital-filtering effect converts the otherwise topologically trivial freestanding Bi lattice into a nontrivial phase; while the reverse is true for Au lattice. The underlying physical mechanism is generally applicable, opening a new and exciting avenue for exploration of large-gap topological surface/interface states.

## Significance

Quantum phase of matter is of great scientific and technological interest. Quantum spin Hall (QSH) insulator is a newly discovered two-dimensional material that exhibits topological edge state residing inside bulk energy gap, so that its edge is metallic with quantized conductance while its bulk is insulating. For its potential applications in spintronics and quantum computing, a large energy gap is desirable, e.g. for room-temperature application. So far, large-gap QSH insulators have been predicted only in freestanding films. Here we demonstrate formation of large-gap QSH state on a semiconductor substrate through epitaxial growth of heavy metal atoms on halogenated Si surface. Our findings not only reveal a new formation mechanism of large-gap QSH insulator, but may also pave the way for its experimental realization.

Topological insulators (TIs) [1–3] are distinguished from conventional insulators by robust metallic surface or edge states residing inside an insulating bulk gap. As these topological states are protected by time reversal symmetry, they have negligible elastic scattering and Anderson localization [4, 5], rendering significant implications in electronic/spintronic and quantum computing devices. In this regard, 2D TI [i.e., quantum spin Hall (QSH) insulator] bears advantage over its 3D counterpart as the edge states of QSH insulator are more robust against nonmagnetic scattering since the only available backscattering channel is forbidden. Many QSH insulators have been discovered [6-20] and most of them have a small energy gap. Recently, there has been an intensive search for 2D TIs with a large energy gap [17-20], which is of both scientific and practical interest, such as for room temperature applications. So far, however, most studied systems rely on freestanding films, and the existence of some 2D freestanding films could be in doubt because of their poor thermal or chemical stability; and even if they do exist, growth and synthesis of freestanding film is usually much harder than growth of thin film on substrate. Furthermore, it is often required to place the functional film on a substrate in a device setting, but the electronic and topological properties of freestanding films will likely be affected by the substrate [21-23]. Therefore, it is highly desirable to search for large-gap QSH states existing on a substrate while maintaining a large gap.

Here, we predict an interesting phenomenon of formation of QSH state on a conventional semiconductor surface with an energy gap as large as ~ 0.8 eV, when heavy metal elements, such as Bi with large spin-orbit coupling (SOC), are grown on 1/3 monolayer (ML) halogen-covered Si(111) surface. Specifically, Si(111) surface functionalized with 1/3 ML of Cl, Br, or I exhibits a Si(111)-$\sqrt{3}\times\sqrt{3}$-X (X=Cl, Br, I) reconstruction of trigonal symmetry due to strong steric repulsion between the halogen atoms, as observed in experiments [24-28]; while F tends to form clusters. The reconstructed Si(111)-$\sqrt{3}\times\sqrt{3}$-X surface provides an ideal template for epitaxial growth of Bi, which self-assembles into a hexagonal superstructure with high thermodynamic stability. Most remarkably, we found that this hexagonal Bi overlayer is 'atomically' bonded to but 'electronically' decoupled from the underlying Si substrate, exhibiting large-gap QSH states completely isolated from Si valence and conduction bands. It originates from an intriguing substrate orbital filtering effect in which the Si(111) substrate effectively selects suitable orbital composition around the Fermi level, to convert the otherwise topological trivial freestanding Bi lattice into a nontrivial substrate-supported QSH insulator.

We have performed density functional theory (DFT) based first-principles calculations (see details in Supplemental Information) of geometry, band structure and band topology of 2D hexagonal lattices of Bi on Si(111)-$\sqrt{3}\times\sqrt{3}$-X (Cl, Br, I) surface. To better illustrate the substrate orbital filtering effect, we also performed similar calculations for hexagonal lattice of Au for comparison. It is worth noting that the Si(111)-$\sqrt{3}\times\sqrt{3}$-X surface has the exposed Si atoms with one dangling bond arranged in hexagonal symmetry (Fig. 1), which serve as the preferred adsorption sites for Bi (Au) atoms, and hence driving the Bi atoms naturally to form a hexagonal lattice. Importantly, the resulting hexagonal Bi lattice is stabilized against segregation to form clusters by the halogen atoms in between the Bi atoms, overcoming a potential problem encountered by direct deposition of metal atoms on clean surface, e.g. deposition of heavy atoms on a magnetic insulator substrate to form Chern insulator [29].

There exists a very strong binding between the deposited Bi and the exposed Si in surface. The calculated adsorption length (*d*) and adsorption energy ($E_{ad}$) of Bi on the Si(111)-$\sqrt{3}\times\sqrt{3}$-Cl surface are found to be 2.46 Å and 2.99 eV, respectively, typical of a covalent bond. Here, $E_{ad} = E_{Bi\,@Cl\text{-}Si(111)} - (E_{Bi} + E_{Cl\text{-}Si(111)})$, where $E_{Bi@Cl-Si(111)}$, $E_{Bi}$ and $E_{Cl\text{-}Si(111)}$ denote the energy of Bi adsorbed Si(111) surface [Bi@Cl-Si(111)], Bi atom, and surface without Bi, respectively. It is also found that forming Bi clusters on the surface is energetically unfavorable; e.g., forming Bi dimer is about 2.5 eV per unit cell higher in energy than the ground state. Moreover, significant energy barrier exists for Bi to diffuse out of the adsorbed site. For instance, an energy barrier of 4.9 eV (5.4 eV) must be overcome for Bi to jump from Si to neighboring Bi (Cl) sites. These results indicate high thermodynamic as well as kinetic stability of the hexagonal Bi overlayer structure.

To examine band topology of Bi@Cl-Si(111) surface, we first purposely exclude SOC from calculation. The band structure of Bi@Cl-Si(111) is shown in Fig. 2(a), along with that of Au@Cl-Si(111) in Fig. 2 (b). In Fig. 2(a), there are two Dirac bands residing inside the bulk gap of Si with a Dirac point at $K$ point, which locates exactly at the Fermi level. Analysis of band composition further showed that the two Dirac bands mainly consist of $p_x$ orbital of Bi. Additionally, there is one weakly dispersive band, consisting of Bi $p_y$ orbital and sitting below the bulk conduction band edge of Si, touches the upper Dirac band at $\Gamma$ point. Another weakly dispersive band, which is a mixture of Bi $p_y$ orbital and the valence bands of Si, touches the lower Dirac band at $\Gamma$ point. We notice that if these two weakly dispersive bands were flat, such type of four-band structure would be ($p_x$, $p_y$) analogue of graphene [30]. In contrast, the band structure of Au@Cl-Si(111) is very different [Fig. 2(b)]. There is no Dirac band or Dirac point; it is a typical semiconductor surface with a large band gap of ~1.0 eV. Its surface states, consisting of mainly 6$s$ orbital of Au, are above the bulk conduction band edge of Si.

Next, the band structures with SOC are shown in Figs. 2(c-d). Comparing Fig. 2(c) with 2(a), one sees that for Bi@Cl-Si(111), two Dirac bands are split apart and a large energy gap of 0.78 eV is opened at $K$ point. Due to band dispersion, the global gap is slightly smaller, ~0.75 eV between the minimum of upper Dirac band somewhere between $\varGamma$ and $K$ point and the maximum of the lower Dirac band at $\varGamma$. Considerable energy gaps are also opened by SOC between the weakly dispersive $p_y$ bands and the Dirac $p_x$ bands. We note that spin degeneracy of these bands is lifted (most noticeable at $K$ point) due to Rashba effect [31] induced by broken inversion symmetry of surface. Again in sharp contrast to the case of Bi, the SOC causes little change in band structure of Au@Cl-Si(111) [comparing Fig. 2(d) with (b)], except some Rashba-type spin splitting. Because standard DFT is known to underestimate semiconductor band gap, we further checked the results with higher level DFT method using screened hybrid functional of Heyd, Scuseria, and Ernzerhof (HSE06, Ref. [32]) and the same electronic behaviors were obtained (see Fig. S1). We also calculated band structures of Bi@Br-Si(111) and Bi@I-Si(111) surfaces (Fig. S2), which show the same physical behavior as Bi@Cl-Si(111) except a smaller gap.

To reveal surface topological properties, we calculated $Z_2$ topology number. As the spatial inversion symmetry is broken, we used a general approach for calculating $Z_2$ by considering the Berry gauge potential and Berry curvature associated with the Bloch wave functions, which does not require any specific point-group symmetry [33, 34]. Indeed, we found that $Z_2 =1$ for Bi@Cl-Si(111), Bi@Br-Si(111) and Bi@I-Si(111) surfaces, confirming their existence of QSH state. In contrast, $Z_2 =0$ for Au@Cl-Si(111). Furthermore, we calculated the topological edge states by constructing edge Green's function of a semi-infinite Bi@Cl-Si(111) surface. The local density of states (DOS) of Bi edge is shown in the inset of Fig. 2(c), which clearly shows gapless edge states connecting the upper and lower bulk band edge to form a 1D Dirac cone at the center of Brillouin zone ($\varGamma$ point). In contrast, no such edge state exists in Au@Cl-Si(111) surface [see inset of Fig. 2(d)].

To understand the physical origin of QSH state in Bi@Cl-Si(111) but not in Au@Cl-Si(111),

we next do an orbital analysis around the Fermi level. Figure 3(a) shows the partial DOS of Bi@Cl-Si(111). It is seen that the $p_z$ orbital of Bi hybridizes strongly with the dangling bond of the exposed surface Si atom overlapping in the same energy range. We calculated the maximally localized Wannier functions (WFs) of Bi by fitting the DFT band structures of a 'hypothetical' freestanding hexagonal lattice with the WANNIER90 code [35]. The resulting WFs are plotted in Fig. 3(b), which show exactly the chemical characteristics of one $s$ and three $p$ orbitals of Bi. While the $s$ orbital of Bi lies in deep energy, interaction with Si effectively removes the $p_z$ bands away from the Fermi level, leaving only the $p_x$ and $p_y$ orbitals [Fig. 3(c)] to form two Dirac bands and two flat bands, which can be described by a four-band model of topological phase in a hexagonal lattice [12, 30]. This indicates that the Cl-Si(111) substrate acts like an orbital filter, to selectively remove the $p_z$ orbitals from the Bi lattice, reducing it from a trivial six-band lattice to a nontrivial four-band lattice, as we explain below.

The concept of 2D TI was originally proposed by Kane and Mele [6] using the graphene model, in which an energy gap is opened at Dirac point in proportion to the strength of SOC. Unfortunately, the SOC of graphene is negligibly small, and much effort has later been devoted to remedy this problem [9, 19]. There are two apparent conditions in the Kane-Mele model to create a 2D TI. One is the lattice symmetry, such as the hexagonal symmetry that produces Dirac band; the other is the SOC. Given these two conditions alone, one might think to create a large-gap 2D TI by constructing a planar hexagonal lattice of heavy metal atoms with large SOC. However, this turned out not to be generally true because there is a third condition of orbital selection associated with the Kane-Mele model.

We have performed first-principles calculations of band structure and band topology of freestanding 2D planar hexagonal lattices of Bi and Au, to examine whether they are 'theoretically' 2D TIs having a large energy gap (See Sec. IV in Supplemental Information). It is found that the two lattices have drastically different electronic and topological properties. The planar Bi lattice is a trivial insulator with $Z_2=0$, while the Au lattice is nontrivial with $Z_2=1$. Their topological difference is originated from the different orbital composition around the Fermi level. For Bi, the valence bands consist of three ($p_x$, $p_y$, and $p_z$) orbitals. The topology associated with the two bands from $p_z$ orbital can be described by the single-orbital two-band Kane-Mele model; while the topology associated with other four bands from $p_x$ and $p_y$ orbitals can be described by the four-band model [12, 30]. Note that separately either the two-band or four-band model gives rise to nontrivial band topology ($Z_2=1$); however, counting all six bands together, the total band topology is trivial ($Z_2=0$), as two odd topological numbers add to an even number. For Au, in contrast, the valence bands mainly consist of single $s$ orbital, and SOC opens a gap of ~70 meV, transforming the lattice into a 2D TI phase. Thus, the planar Au lattice can also be understood by the Kane-Mele model, except that it involves a single $s$ orbital rather than the $p_z$ orbital in graphene.

To transform the topological trivial planar hexagonal lattice of Bi into a nontrivial phase, one efficient way is to select one $p_z$ or two $p_x$ and $p_y$ orbitals, to realize the two-band or four-band model. In our Bi@Si(111)-$\sqrt{3} \times \sqrt{3}$-X (Cl, Br, I) system, the exposed Si atom in the Cl-Si(111) surface interacts strongly with Bi and removes the $p_z$ orbital of Bi, with $p_x$ and $p_y$ orbitals remain active near Fermi energy, as shown in Fig. 3(a). In contrast, the Au atom has single $s$-orbital valence electron, and when it is bonded with the exposed surface Si atom, it simply saturates the Si dangling bond just like the Cl does. Consequently, the band structure of Au@Cl-Si(111) is essentially the same as that of the semiconducting pristine Cl-Si(111) surface. Specifically, we can describe the Bi@Cl-Si(111) using a simplified ($p_x$, $p_y$) four-band model Hamiltonian in a hexagonal lattice [12, 30], which produces either a flat-band Chern insulator, or a 2D TI, depending on the location of the Fermi level. Due to large SOC of Bi (~1.25 eV), QSH state with an energy gap as large as ~0.8 eV is found, which is possibly the largest gap so far predicted.

We believe that experimentally, it is highly feasible to realize the large-gap QSH states as we propose here, based on the existing related experiments. First, the required template of halogenated Si surface has already been widely studied in early surface science research [24-26], especially the 1/3 ML halogen-covered Si(111) surface exhibits the exact trigonal-symmetry reconstruction [25, 27, 28]. Second, the epitaxial growth of metal overlayer on the 1/3 ML halogen-Si(111) template should be highly possible. For instance, hexagonal lattices of indium overlayer has been successfully grown on the Si(111) $\sqrt{3} \times \sqrt{3}$ -Au surface [36], although the surface bands of In hybridize strongly with those of underlying Si substrate [37], making this system topologically trivial. Nevertheless, the growth process can be borrowed for our purposes in realizing QSH states. Furthermore, it should be possible to grow stripes of Bi hexagonal lattice on the halogenated Si(111) surface, so that the helical edge state is naturally created for measurement at the domain boundary of the fully chlorinated Si(111) surface (a trivial insulator or conventional semiconductor) and the Bi@Cl-Si(111) stripe (a QSH insulator) [20]. We envision that realization of topological edge states on Si surface, which may be used as dissipationless interconnects for electronic circuits [38], to significantly advance the current Si-based technology.

## Methods

Our electronic structure calculations based on DFT were performed by using a plane wave basis set, the projector-augmented wave method and the Perdew-Burke-Ernzerhof generalized gradient approximation for exchange-correlation potentials, as implemented in the Vienna *ab* *initio* package (VASP) code [39]. Details are presented in Supplemental Information.

## Acknowledgements

This research was supported by DOE (Grant No: DEFG02-04ER46148); Z. F. Wang and W. Ming additionally thank support from NSF-MRSEC (Grant No. DMR-1121252). We thank NERSC and the CHPC at University of Utah for providing the computing resources.

## Footnotes

Author contributions: M. Z. carried out the theoretical calculations with the assistance of W. M. M., Z. L., Z. F. W. and P. L.; F. L. guided the overall project; M. Z. and F. L. wrote the manuscript.

Supplementary Information accompanies this paper is available.

Competing financial interests: The authors declare no conflict of interest.

## References

[1] Moore JE (2010) The birth of topological insulators. *Nature* 464: 194−198.

[2] Hasan MZ, Kane CL (2010) Colloquium: topological insulators. *Rev Mod Phys* 82: 3045−3067.

[3] Qi XL, Zhang SC (2011) Topological insulators and superconductors. *Rev Mod Phys* 83: 1057−1110.

[4] Bernevig BA, Zhang, SC (2006) Quantum spin hall effect. *Phys Rev Lett* 96: 106802.

[5] Xu C, Moore JE (2006) Stability of the quantum spin Hall effect: Effects of interactions, disorder, and $Z_2$ topology. *Phys Rev B* 73: 045322.

[6] Kane CL, Mele EJ (2005) Quantum spin Hall effect in graphene. *Phys Rev Lett* 95: 226801.

[7] Bernevig BA, Hughes TL, Zhang SC (2006) Quantum Spin Hall Effect and Topological Phase Transition in HgTe Quantum Wells. *Science* 314: 1757−1761.

[8] König M, et al. (2006) Quantum spin Hall insulator state in HgTe quantum wells. *Science* 318: 766–770.

[9] Weeks C, et al. (2011) Engineering a robust quantum spin Hall state in graphene via adatom deposition. *Phys Rev X* 1: 021001.

[10] Liu CC, Feng W, Yao Y (2011) Quantum spin Hall effect in silicene and two-dimensional germanium. *Phys Rev Lett* 107: 076802.

[11] Wang ZF, Liu Z, Liu F (2013) Organic topological insulators in organometallic lattices. *Nat Commun* 4: 1471–1475.

[12] Liu Z, et al. (2013) Flat Chern Band in a Two-Dimensional Organometallic Framework. *Phys Rev Lett* 110: 106804.

[13] Wang ZF, Su N, Liu F (2013) Prediction of a Two-Dimensional Organic Topological Insulator. *Nano Lett* 13: 2842–2845.

[14] Wang ZF, Liu Z, Liu F (2013) Quantum anomalous Hall effect in 2D organic topological insulator. *Phys Rev Lett* 110: 196801.

[15] Murakami S (2006) Quantum spin Hall effect and enhanced magnetic response by spin-orbit coupling. *Phys Rev Lett* 97: 236805.

[16] Wada M, Murakami S, Freimuth F, Bihlmayer G (2011) Localized edge states in two-dimensional topological insulators: Ultrathin Bi films. *Phys Rev B* 83: 121310.

[17] Liu Z, et al. (2011) Stable nontrivial $Z_2$ topology in ultrathin Bi (111) films: a first-principles study. *Phys Rev Lett* 107: 136805.

[18] Zhang PF, et al. (2012) Topological and electronic transitions in a Sb(111) nanofilm: The interplay between quantum confinement and surface effect. *Phys Rev B* 85: 201410.

[19] Hu J, Alicea J, Wu R, Franz M (2012) Giant topological insulator gap in graphene with 5d adatoms. *Phys Rev Lett* 109: 266801.

[20] Xu Y, et al. (2013) Large-Gap Quantum Spin Hall insulators in Tin Films. *Phys Rev Lett* 111: 136804.

[21] Yang F, et al. (2012) Spatial and energy distribution of topological edge states in single Bi(111) bilayer. *Phys Rev Lett* 109: 016801.

[22] Hirahara T, et al. (2012) Atomic and electronic structure of ultrathin Bi(111) films grown on $Bi_2Te_3$(111) substrate: Evidence for a strain-induced topological phase transition. *Phys Rev Lett* 109: 227401.

[23] Wang ZF, et al. (2013) Creation of helical Dirac fermions by interfacing two gapped systems of ordinary fermions. *Nat Commun* 4: 1384−1839.

[24] Buriak JM (2002) Organometallic chemistry on silicon and germanium surfaces. *Chem Rev* 102: 1271–1308.

[25] Dev BN, Aristov V, Hertel N, Thundat T, Gibson, WM (1985) Chemisorption of Bromine on cleaved silicon (111) surface: An X-ray standing wave interference spectrometric analysis. *Surf Sci* 163: 457.

[26] Rivillon S, et al. (2005) Chlorination of hydrogen-terminated silicon (111) surfaces. *J Vac Sci Technol A* 23: 1100.

[27] Ferguson GA, Rivillon S, Chabal Y, Raghavachari K (2009) The structure and vibrational spectrum of the Si(111)-H/Cl surface. *J Phys Chem C* 113: 21713–21720.

[28] Michalak DJ, et al. (2010) Nanopatterning Si(111) surfaces as a selective surface-chemistry route. *Nat Mater* 9: 266–271.

[29] Garrity KF, Vanderbilt D (2013) Chern insulators from heavy atoms on magnetic substrates. *Phys Rev Lett* 110: 116802.

[30] Wu C, Bergman D, Balents L, Das Sarma S (2007) Flat Bands and Wigner Crystallization in the Honeycomb Optical Lattice. *Phys Rev Lett* 99: 070401.

[31] Bychkov YA, Rashba, EI (1984) Properties of a 2D electron gas with lifted spectral degeneracy. *JETP Lett* 39: 78–81.

[32] Heyd J, Scuseria GE, Ernzerhof M (2003) Hybrid functionals based on a screened Coulomb potential. *J Chem Phys* 118: 8207.

[33] Feng W, et al. (2012) First-principles calculation of $Z_2$ topological invariants within the FP-LAPW formalism. *Comput Phys Commun* 183: 1849–1859.

[34] Xiao D, et al. (2010) Half-Heusler Compounds as a New Class of Three-Dimensional Topological Insulators. *Phys Rev Lett* 105: 096404.

[35] Mostofi AA, et al. (2008) Wannier90: a tool for obtaining maximally-localized wannier functions. *Comput Phys Commun* 178: 685–699.

[36] Gruznev DV, et al. (2006) Si(111)-α-$\sqrt{3} \times \sqrt{3}$-Au phase modified by In adsorption: Stabilization of a homogeneous surface by stress relief. *Phys Rev B* 73: 115335.

[37] Hsu CH, Lin WH, Ozolins V, Chuang FC (2012) Electronic structure of the indium-adsorbed Au/Si(111)-$\sqrt{3} \times \sqrt{3}$ surface: A first-principles study. *Phys Rev B* 85: 155401.

[38] Zhang X, Zhang SC (2012) Chiral interconnects based on topological insulators. *Proc SPIE Int Soc Opt Eng* 8373: 837309.

[39] Kresse G, Furthmuller J (1996) Efficient iterative schemes for ab initio total-energy calculations using a plane-wave basis set. *Phys Rev B* 54: 11169.

**Figure Legends**

**Fig. 1. Schematic illustration of the epitaxial growth of large-gap QSH states on Si substrate.** We propose to fabricate the hexagonal lattices of heavy metal by direct deposition of heavy metal (HM) atoms onto Si(111)-$\sqrt{3}\times\sqrt{3}$-X (X=Cl, Br, I) template. The surface unit cell vector ($\boldsymbol{a_1}, \boldsymbol{a_2}$) is also indicated.

**Fig. 2. Band structures of Bi and Au on Si(111)-$\sqrt{3}\times\sqrt{3}$-Cl surface.** (a) and (b) for Bi and Au without SOC, respectively. (c) and (d) with SOC. Bands compositions near Fermi level are indicated. Inset of (c) shows the Dirac edge states within the SOC-induced band gap in Bi@Cl-Si(111), and inset of (d) shows no edge states within the band gap of Au@Cl-Si(111). The brightness is proportional to the magnitude of the states.

**Fig. 3. Physical origin of QSH states on Si substrate.** (a) The partial DOS projected onto $p_x$, $p_y$, and $p_z$ orbitals of Bi, and the total DOS of neighboring Si atoms for Bi@Cl-Si(111). (b) The calculated Wannier functions characterized by $s$, $p_x$, $p_y$ and $p_z$ orbitals of freestanding hexagonal lattice of Bi, with only $p_x$ and $p_y$ located near Fermi level. (c) Illustration of hexagonal lattice made of $p_x$ and $p_y$ orbitals on each site.

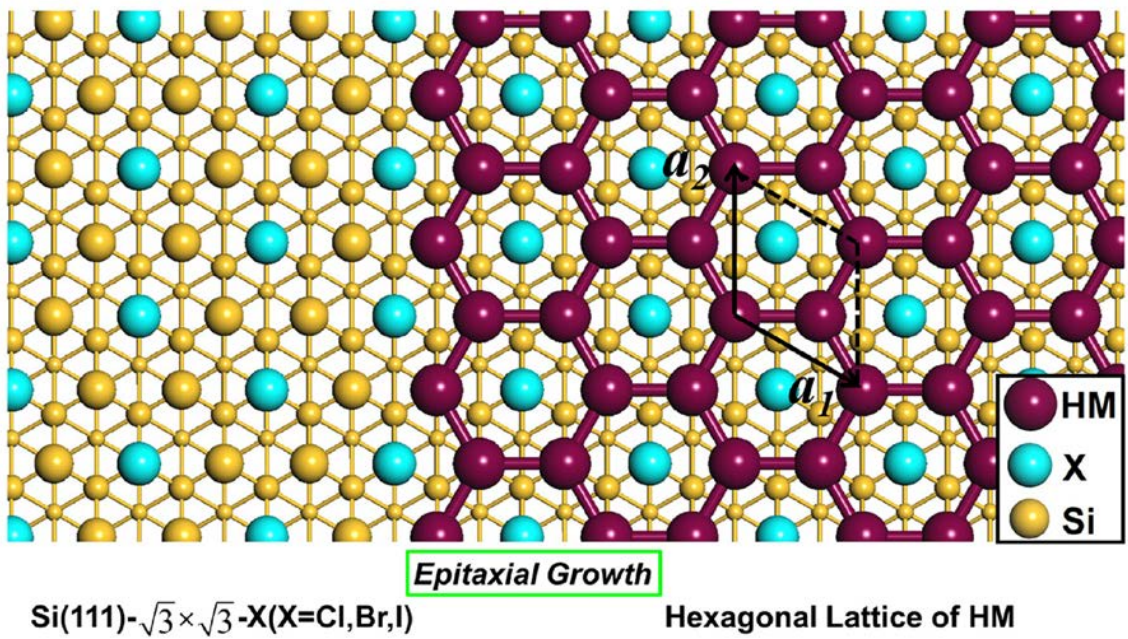

**Fig. 1**

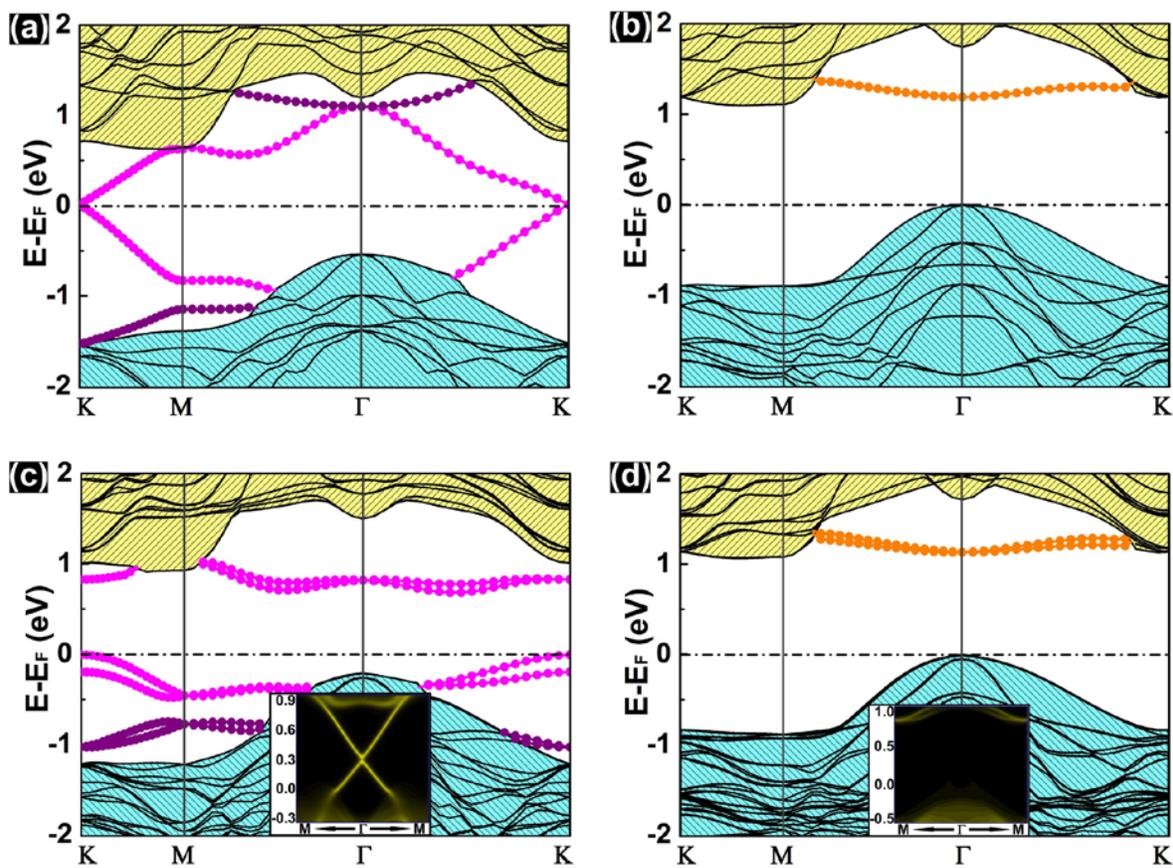

**Fig. 2**

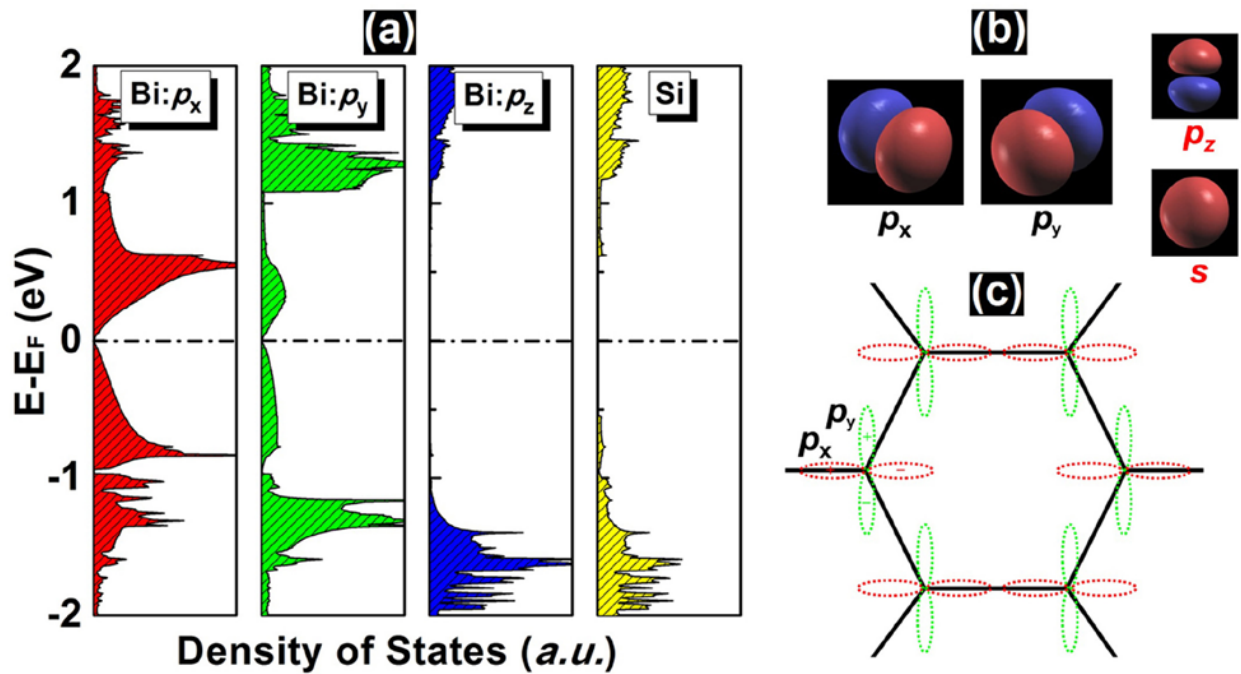

**Fig. 3**

# -Supplemental Information-

# Epitaxial growth of large-gap quantum spin Hall insulator on semiconductor surface: The substrate orbital filtering effect

*Miao Zhou[1], Wenmei Ming[1], Zheng Liu[1], Zhengfei Wang[1], Ping Li[1,2], and Feng Liu[1,3*]*

[1] Department of Materials Science and Engineering, University of Utah, UT 84112

[2] School of Physics and Technology, University of Jinan, Jinan, Shangdong, China 250022

[3] Collaborative Innovation Center of Quantum Matter, Beijing, China 100871

*Corresponding author:

Address: Room 304, 122 S. Central Campus Drive, Salt Lake City, UT 84112

Tel: 801-587-7719

Email: fliu@eng.utah.edu

**CONTENTS**

## I. Models and computational details

First-principles electronic structure calculations based on density functional theory (DFT) were carried out using the plane-wave-basis-set and the projector-augmented-wave method [1, 2], as implemented in the VASP code [3]. The energy cutoff was set to 500 eV. For the exchange and correlation functional, the generalized gradient approximation (GGA) in Perdew-Burke-Ernzerhof (PBE) format [4] was used. The calculation was further verified by using the more sophisticated Heyd-Scuseria-Ernzerhof hybrid functional (HSE06, Ref. [5]), which made no essential difference (see Fig. S1). Spin-orbit coupling (SOC) is included by a second variational procedure on a fully self-consistent basis.

Si(111) surfaces were modeled by using a slab geometry of ten atomic layers, with a vacuum region of 30 Å in the direction normal to the surface. Test calculations were performed by using larger thickness (twelve and sixteen layers) which gave similar results. The bottom Si surfaces were terminated by H atoms in a monohydride form. During structural optimization, both the tenth layer of Si atoms and the H atoms saturating them were fixed and all other atoms were fully relaxed until the atomic forces are smaller than 0.01 eV/Å. A 15×15×1 $\Gamma$-centered $k$-point mesh was used to sample the Brillouin zone. Dipole corrections were also tested and found making little difference.

For epitaxial growth of Bi and Au atoms on the 1/3 X-covered Si(111) surface (X=Cl, Br, I), we considered a $\sqrt{3}\times\sqrt{3}$ supercell and different adsorption configurations. The energy barriers for Bi hopping on the Si(111) surfaces were calculated by using the climbing-image nudged elastic band method [6].

$Z_2$ invariant calculations were performed by using the full-potential linearized augmented plane-wave method [7] within the GGA-PBE functional including SOC. A converged ground state was obtained using 5000 k-points in the first Brillouin zone and $K_{max} \times R_{MT} = 8.0$, where $K_{max}$ is the maximum size of the reciprocal lattice vectors and $R_{MT}$ denotes the muffin-tin radius. Wave functions and potentials inside the atomic sphere are expanded in spherical harmonics up to $l = 10$ and 4, respectively. For $Z_2$ calculation, we follow the method by Fukui *et al.* [8], to directly perform the lattice computation of the $Z_2$ invariants from first-principles.

## II. Results by using HSE06 hybrid functional

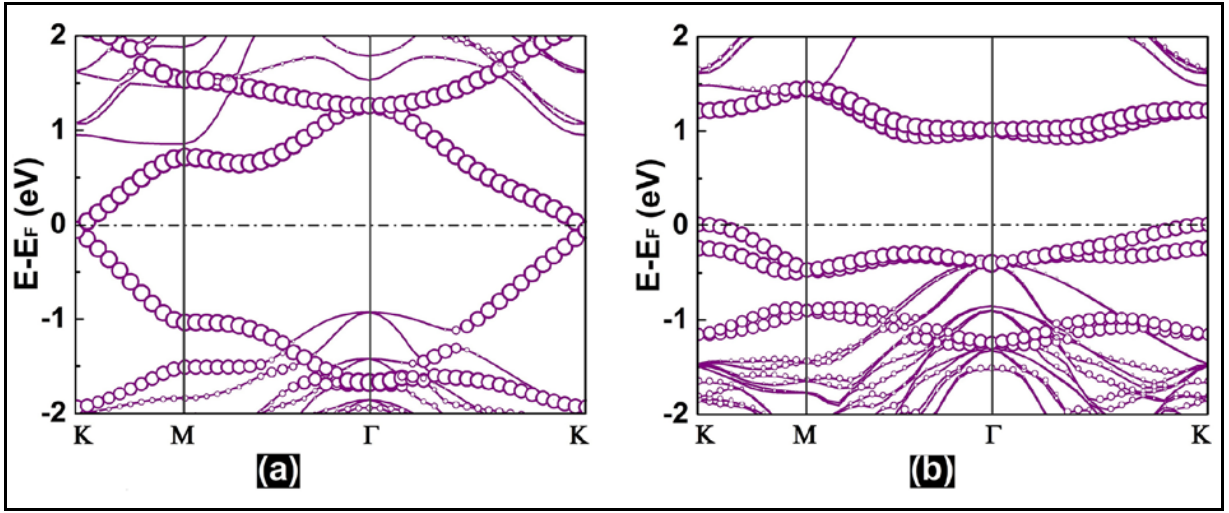

**Fig. S1. Band structures for Bi@Cl-Si(111) surface system calculated by using HSE06 hybrid function.** (a) Without SOC and (b) with SOC. Bands compositions are indicted, with size of circles denoting the contribution from Bi.

## III. Band structures of Bi@Br(I)-Si(111)

We also calculated band structures of hexagonal Bi lattice grown on Si(111)-$\sqrt{3}\times\sqrt{3}$-Br and -I surfaces without and with SOC, as shown in Fig. S2. They show similar electronic properties and SOC-induced band gap opening as those of Bi@Cl-Si(111) [see Figs. 2 (a) and (c) in the paper].

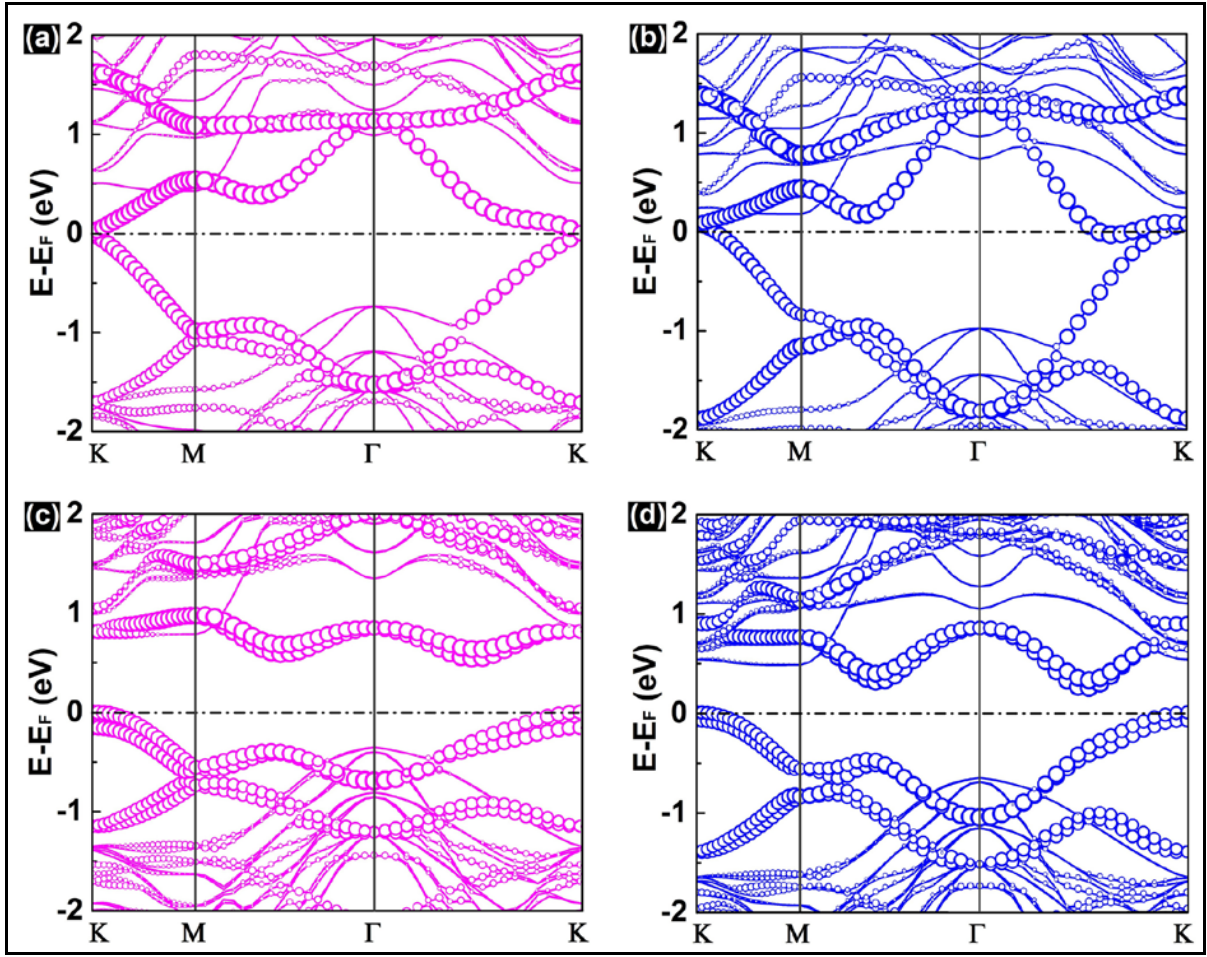

**Fig. S2. Band structures for Bi@Br (I)-Si(111).** (a) and (b) Band structures for Bi@Br-Si(111) and Bi@I-Si(111) surface structure without SOC, respectively. (c) and (d) Same as (a) and (b) with SOC. Bands compositions are indicted in the same way as in Fig. S1.

It is interesting to note that the Bi-related energy bands become more dispersive for Bi@Br-Si(111) and Bi@I-Si(111) surface. This results from different interaction strength between Bi atoms mediated by different halogen ions on the surface. Going from Cl to Br and to I, the ionic radius gradually increases from 1.81 Å to 1.99 Å and to 2.16 Å, respectively. Consequently, the Bi-Bi interaction mediated by larger halogen ions is stronger with larger orbital overlap, leading to larger hopping and more dispersive bands. This in turn leads to smaller SOC-induced energy gap for Bi@Br-Si(111) (0.6 eV ) and Bi@I-Si(111) (0.4 eV) [Figs.

S2(c) and (d)]. In general, the Bi-Bi interaction can also be tuned by choosing different semiconductor substrates, which is an interesting topic for future study.

**IV. Electronic and topological properties of free-standing planar Bi/Au hexagonal lattice**

Figures S3(a) and (b) show the calculated band structures and DOS around Fermi level of Bi and Au, respectively. Surprisingly, the two lattices are found to have drastically different electronic and topological properties. The planar Bi lattice is a trivial insulator with $Z_2=0$, while the Au lattice is nontrivial with $Z_2=1$. Their topological difference is found to originate from the different orbital composition around the Fermi level. For Bi, the valence bands consist of three ($p_x$, $p_y$, and $p_z$) orbitals [Figs. S3(a)]. The topology associated with the two bands from $p_z$ orbital can be described by the single-orbital two-band Kane-Mele model; while the topology associated with other four bands from $p_x$ and $p_y$ orbitals can be described by the four-band model (Refs. [12, 30] in the paper]. Note that separately either the two-band or four-band model gives rise to nontrivial band topology ($Z_2=1$); however, counting all six bands together, the total band topology is trivial ($Z_2=0$), as two odd topological numbers add to an even number. For Au, in contrast, the valence bands mainly consist of single $s$ orbital, which form a Dirac point at $K$ point without SOC, as shown in Fig. S3(b). The SOC opens a gap of ~70 meV, transforming the lattice into a 2D TI phase. Thus, the planar Au lattice can also be understood by the Kane-Mele model, except that it involves a single $s$ orbital rather than the $p_z$ orbital in graphene.

More generally, we can better understand the topological phases in a 2D hexagonal lattice by a multi-orbital tight-binding model. The effective Hamiltonian with the nearest-neighbor hopping and intrinsic SOC can be written as,

$$\hat{H} = \sum_{i,l} \varepsilon_l c_{il}^\dagger c_{il} - \sum_{<i,j>l,l'} (t_{ijll'} \cdot c_{il}^\dagger c_{jl'} + H.c.) + \lambda_{so} \hat{L} \cdot \hat{\sigma} ,  \quad (1)$$

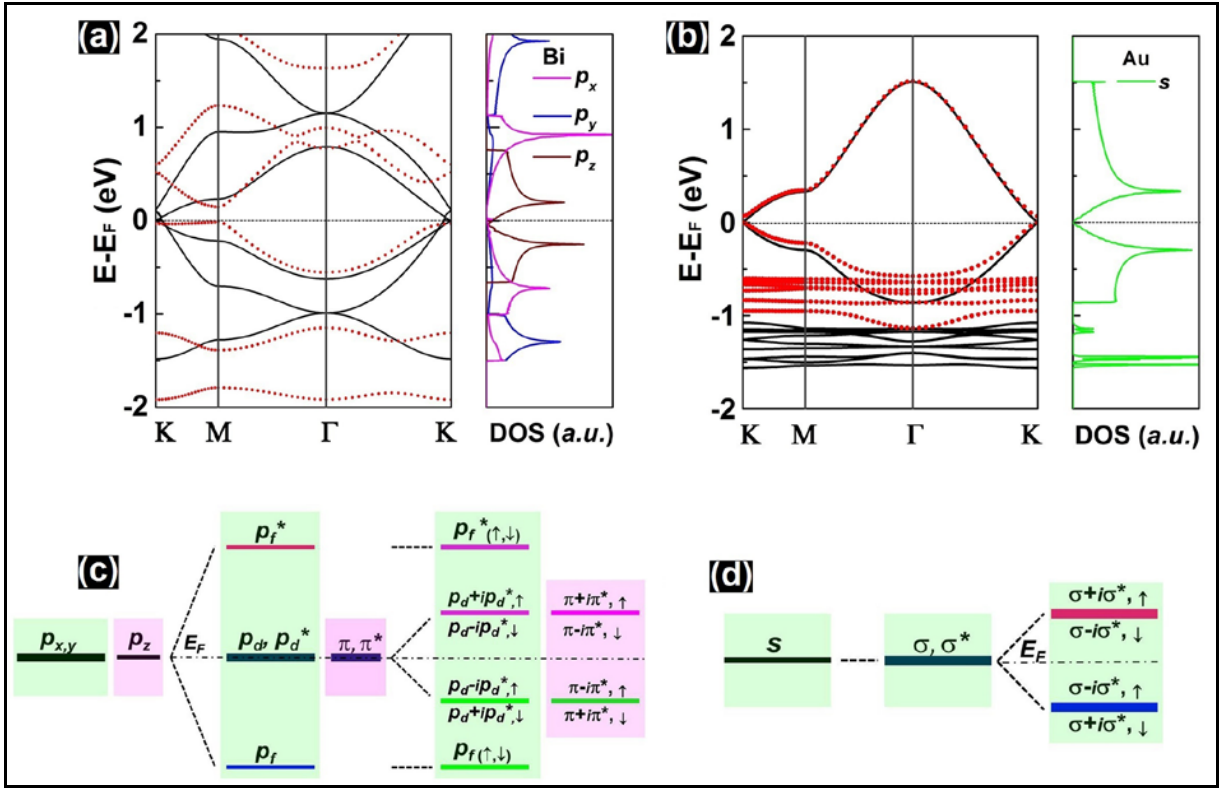

**Fig. S3. Band structure, DOS and energy diagram of planar Bi and Au hexagonal lattice.** (a) Band structure without (black solid curves) and with SOC (red dotted curves) of a planar hexagonal lattice of Bi, along with the atomic-orbital projected DOS without SOC. (b) Same as (a) for Au. (c) and (d) The energy diagrams of the Bi and Au lattices at *K* point illustrating the effects of orbital hybridization and SOC.

where $c_{il}^{\dagger}$ creates an electron of *l*-th orbital ($s$, $p_x$, $p_y$, $p_z$, $d_{xy}$, ...) at site *i*. $\varepsilon_l$ ($t_{ijll'}$) denotes the on-site energy (hopping energy) and $\lambda_{SO}$ is the strength of SOC. As mentioned above, the Bi lattice can be described by the *p*-orbital six-band model, as illustrated in Fig. S3(c). Due to the planar symmetry, $p_x$ and $p_y$ oribitals hybridize to be distinguished from the $p_z$ orbital, resulting in two branches of energy bands. The $p_z$ branch of $\pi$ and $\pi^*$ bands is exactly the same as graphene (Ref. [6] in the paper). The ($p_x$, $p_y$) branch has four bands: two flat bands ($p_f$, $p_f^*$) bracketing two dispersive bands ($p_d$, $p_d^*$) which form a Dirac point at *K* point. The SOC opens a gap in the dispersive bands at *K* point and mixes the $p_d$ and $p_d^*$ bands into two sets of $p_d \pm i\, p_d^*$ bands
23

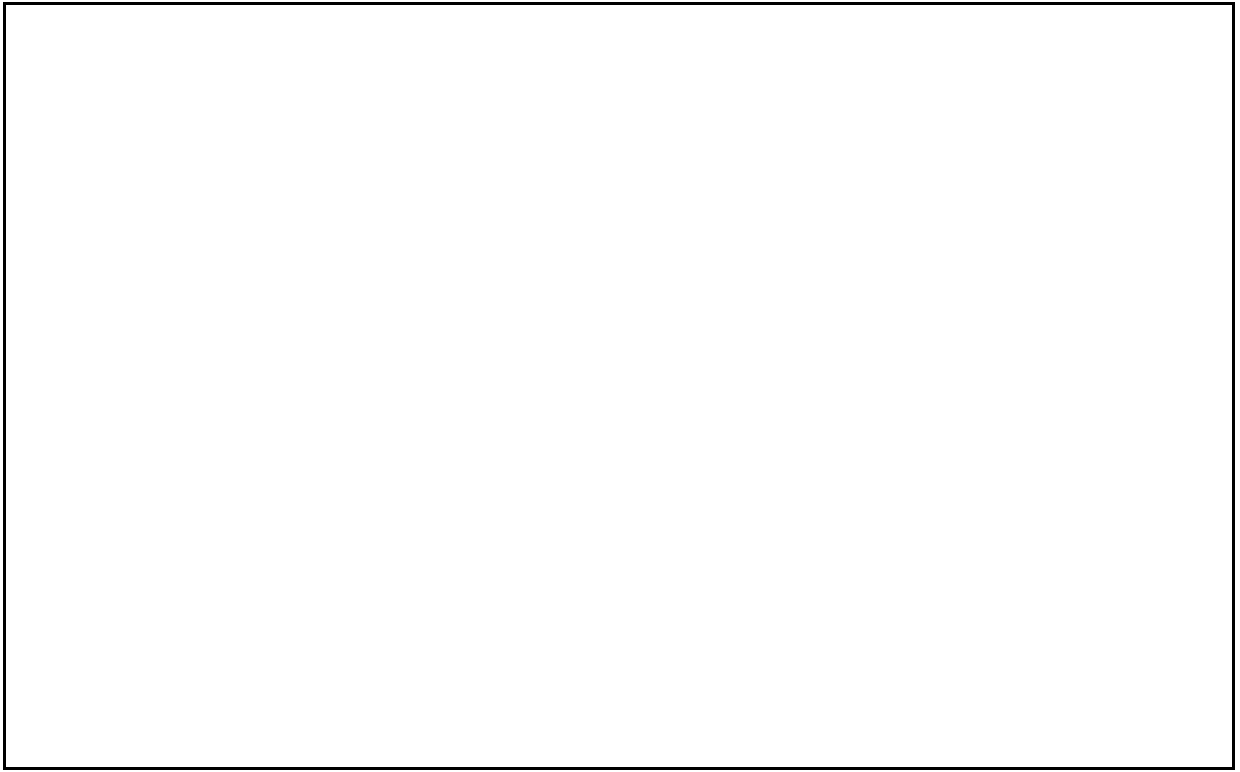

**Fig. S4. Band structure, DOS and energy diagram of buckled and H-saturated Bi hexagonal lattice.** (a-c) Structural model, band structure and atomic-orbital projected DOS, and the energy diagram (at $\Gamma$ point), respectively, of a buckled Bi(111) bilayer. The band inversion is highlighted by a dashed rectangle. (d-f) Same as (a-c) for planar Bi hexagonal lattice with one side saturated by H.

encoding a nontrivial topology. Again, the two branches of bands are both topologically nontrivial, but their sum becomes trivial. The planar Au lattice can be described by the $s$-orbital two-band model, as illustrated in Fig. S3(d). Without SOC, the two $s$ orbitals hybridize into linearly dispersive two-fold degenerate $\sigma$ and $\sigma^*$ bands which touches at $K$ point (Dirac point); the SOC opens a gap and mixes the $\sigma$ and $\sigma^*$ bands into two sets of $\sigma \pm i\sigma^*$ bands encoding a nontrivial topology.

There are two ways to make the planar hexagonal Bi lattice topologically nontrivial. The first way is by the well-known band inversion approach (Ref. [7] in the paper), which in the present case can be achieved by buckling the lattice into a non-planar structure [Fig. S4(a)], i.e., the single Bi(111) bilayer. Figure S4(b) shows the band structure and DOS of a Bi(111) bilayer, which is confirmed with nontrivial topology (Ref. [17] in the paper). In such a buckled structure, the Bi-Bi bond angle is around 90°, indicating that three ($p_x$, $p_y$ and $p_z$) orbitals are degenerate with each other. Chemical bonding and crystal field splitting lifts the degeneracy and form one set of doubly degenerate $\sigma_{1,2}$ and $\sigma_{1,2}*$ bands and another set of non-degenerate $\sigma_3$ and $\sigma_{3*}$ bands, in the order of energy as shown in Fig. S4(c). The SOC opens an energy gap and further lifts the degeneracy of $\sigma_{1,2}$ and $\sigma_{1,2}*$ bands, as well as causes a band inversion of energy order between $\sigma_{1\pm i2}$ and $\sigma_3*$ bands around the Fermi level. Consequently, the overall band topology becomes nontrivial.

The second approach is to simply remove one branch of orbitals [either ($p_x$, $p_y$) or $p_z$] to reduce the trivial six-band lattice into a nontrivial two- or four-band lattice. To verify this idea, we artificially saturate the planar hexagonal Bi lattice with H to remove the $p_z$ orbital [Fig. S4(d)]. It is found that the $p_z$ orbital of Bi hybridizes strongly with $s$ orbital of H, shifting away from the Fermi level, so that the system reduces to a ($p_x$, $p_y$)-orbital four-band model, which supports a nontrivial topological phase [Figs. S4(e) and S4(f)]. This is essentially what happens with the Bi@Cl-Si(111), where the exposed Si atom in the Cl-Si(111) surface fulfills the role of H atom to remove the $p_z$ orbital of Bi.

**Reference**

[1] Blöchl PE (1994) Projector augmented-wave method. *Phys Rev B* 50: 17953.

[2] Kresse G, Joubert D (1999) From ultrasoft pseudopotentials to the projector augmented-wave method. *Phys Rev B* 59: 1758.

[3] Kresse G, Hafner J (1993) *Ab initio* molecular dynamics for liquid metals. *Phys Rev B* 47: 558.

[4] Perdew JP, Burke K, Ernzerhof M (1996) Generalized Gradient Approximation Made Simple. *Phys Rev Lett* 77: 3865.

[5] Heyd J, Scuseria GE, Ernzerhof M (2006) Hybrid functionals based on a screened Coulomb potential. *J Chem Phys* 124: 219906.

[6] Henkelman G, Uberuaga BP, Jónsson H (2000) A climbing image nudged elastic band method for finding saddle points and minimum energy paths. *J Chem Phys* 113: 9901.

[7] Singh DJ, Nordstrom L (1994) Planewaves, Pseudopotentials and the LAPW Method (Kluwer Academic, Boston).

[8] Fukiu T, Hatsugai Y (2007) Quantum Spin Hall Effect in Three Dimensional Materials: Lattice Computation of $Z_2$ Topological Invariants and Its Application to Bi and Sb. *J Phys Soc Jpn* 76: 053702.